\documentclass[pre,twocolumn,aps,eqsecnum]{revtex4}
\usepackage{epsfig}

\begin{document}
\begin{titlepage}

\title{Statistical Thermodynamics of Crystal Plasticity} 
\author{J.S. Langer}
\affiliation{Department of Physics, University of California, Santa Barbara, CA  93106-9530}
\date{\today}
\begin{abstract}
This article is written in memory of Pierre Hohenberg with appreciation for his deep commitment to the basic principles of theoretical physics.  I summarize recent developments in the theory of dislocation-enabled deformation of crystalline solids.  This topic is especially appropriate for the Journal of Statistical Physics because  materials scientists, for decades, have asserted that statistical thermodynamics is not applicable to dislocations.  By use of simple, first-principles analyses and comparisons with experimental data, I argue that these people have been wrong, and that this field should now be revitalized because of its wide-ranging intellectual and technological importance.
\end{abstract}

\maketitle

\end{titlepage}

\section{Pierre}

It's hard for me to understand that Pierre Hohenberg is no longer with us.  We knew each other thoughout all of our professional lives -- never directly collaborating in research -- but often working together on physics-related projects and counting on each other for friendly advice and criticism.  Our last joint project was a memorial for Walter Kohn who had been a prime mentor for both of us, in different ways and in different places, throughout our careers.  

Pierre's scientific style was always more focussed than mine on fundamental principles and rigorous analyses.  His first major success was the famous Hohenberg-Kohn variational theorem that provided a firm starting point for density-functional theory; and his last major effort was his attempt to construct a rigorous basis for quantum mechanics.  In December 2014, Pierre and I were two of the honorary birthday guests at Joel Lebowitz's Statistical Mechanics Conference at Rutgers.  I talked about an utterly non-rigorous (and probably wrong) theory of glass transitions.  As I remember, Pierre sat in the front row, mostly looking grumpy and only occasionally chuckling. This was not his kind of theoretical physics.  

At about the same time, however, I was starting a different adventure that might have amused Pierre had he had the time or inclination to pay attention to it.  This was my still ongoing attempt to reformulate the dislocation theory of crystal plasticity -- a central part of materials science that has remained amazingly undeveloped for more than half a century.  I've been trying to do this in a way that is consistent with the basic principles of statistical physics and that, at the same time, is realistic enough to be tested by experiment.  Perhaps this is just geriatric delusion, but I'm now fairly sure that I've made progress.  

Much of dislocation theory as practiced today is based on questionable and sometimes demonstrably incorrect phenomenological assumptions.  As a result, there are many technologically important behaviors that we have not understood and which need now to be restudied. The new ``thermodynamic dislocation theory'' (TDT), whose outlines I describe here, looks like a good start in this direction.  In memory of Pierre's commitment to basic principles, I devote my remarks here to a description of the TDT based primarily on the second law of thermodynamics and dimensional analysis.  I also emphasize that I have experimental confirmation as well as mathematics to support my heresies.

\section{History}

First -- some history.  It was discovered in the 1930's, notably by G.I. Taylor \cite{TAYLOR-34} and E. Orowan and colleagues, that the deformation of a rigid crystal can occur at the cost of relatively little energy via the motion of crystalline defects known as a dislocations. For simplicity, think of a dislocation as a line that marks the boundary of an extra partial plane of atoms within a crystal. When such a line moves across the interior of the system, the extra plane effectively moves with it, and part of the crystal deforms by about an atomic spacing.  The decades following this insight were devoted largely to studying the properties of individual dislocations, e.g. the  stress needed to move the dislocation through the lattice, and the interactions between dislocations and various crystalline defects.  Standard references from this era include \cite{COTTRELL-53,FRIEDEL-67,HIRTH-LOTHE-68}.

The big problem was work hardening, which is the question of why the stress required for deformation increases as the material is deformed. With the advent of modern microscopy, it became clear that the density of dislocations increases with deformation, so that the dislocation lines become entangled and increasingly immobile.  Cottrell and Nabarro \cite{COTTRELL-02} described this entanglement as a ``birds' nest,''  implying  that it was too complex to be amenable to conventional statistical or thermodynamic analyses. 

In his seminal book on dislocation theory \cite{COTTRELL-53}, Cottrell used an entropic argument to assert that  ``the dislocation cannot exist as a thermodynamically stable lattice defect.'' He was technically correct.  However,  a more accurate conclusion from his argument is that dislocations must be intrinsically nonequilibrium objects.  They are created, annihilated, and enable deformations only in externally driven systems through which energy is flowing.  This is the basic premise of the thermodynamic dislocation theory (TDT) \cite{LBL-10}, which is the main topic of this paper.   But Cottrell's argument has long been interpreted to mean that statistical thermodynamics is irrelevant to dislocation theory.  Fifty years after publication of his book, he stated that ``... the theory is ... still at the stage of merely being interpretative, not predictive.''\cite{COTTRELL-02}  By then, he had been joined in this conclusion by most materials scientists, with the result that basic theoretical research in this field has been at a standstill since the 1950's.  

By calling this situation a ``standstill,'' I certainly do not mean that nothing has  been done.  On the contrary, there has been a large amount of increasingly sophisticated experimental observation of dislocations moving in complex environments.  We have seen images of dislocations interacting with stacking faults, piling up at grain boundaries, forming cellular structures, etc..  The trouble is that, without basic understanding, we cannot know how any of these observations might actually be relevant to large-scale plasticity. 

We also have seen increasingly detailed measurements of stresses as functions of strain, strain rate, and temperature.  To interpret these measurements, theorists have postulated increasingly complicated curve-fitting formulas with arbitrary power laws, thermal activation factors, and the like. (For example, see \cite{KOCKS-MECKING-03,PTW-03}.)  Unfortunately, in writing these formulas, the theorists commonly have assumed that flow stresses associated with different kinds of impediments to dislocation motion contribute independently and additively to the total flow stress.\cite{GRAY-12,ARMSTRONG-HP14}   This additivity assumption has never been justified theoretically.  As shown by a clear counterexample in \cite{JSL-18}, it is not generally correct. Moreover, the curve-fitting efforts have completely ignored the entanglement question that, according to Cottrell, is the crux of the problem.  In short, these efforts are not based on fundamental principles; in Cottrell's words, they are ``not predictive.''

My summary of the state of this field would not be complete without briefly mentioning two other currently active areas of research.  One of these is a first-principles family of continuum theories; the other is numerical simulation. 

The continuum or ``strain-gradient'' theories \cite{FLECKetal-94} use a continuous deformation field as the fundamental dynamical variable, and identify the curl of this field as the local density of dislocations.  In doing this, they necessarily assume a fixed frame of reference from which the deformation is measured.  But plasticity is an irreversible phenomenon that, by definition, loses its memory of any initial reference state; thus, these theories can at best be valid for small quasistatic deformations and cannot tell us much about strain hardening.  

Nevertheless, there is an interesting aspect of these theories. They describe what are known as ``geometrically necessary'' dislocations,  which are the extra dislocations needed in order for crystalline lattices to undergo bending deformations without storing unphysically huge amounts of elastic energy.  The other dislocations are usually called ``redundant'' because they occur in elastically neutral pairs and are invisible in the continuum analyses.  Most of the dislocations in Cottrell's birds' nest are ``redundant;'' they are the entangled dislocations that account for most of the resistance to plastic flow.  Nevertheless, the geometrically necessary dislocations are indeed ``necessary,'' for example, to determine deformations at the tips of notches in fracture-toughness measurements, or to understand how memories of deformation are stored to produce what are called ``Bauschinger effects.''  There is a great deal of interesting work remaining to be done along these lines.  

Numerical simulations play roles somewhere between theory and experiment.  Until quite recently, the most popular of these simulations were based on what is called ``discrete dislocation dynamics,'' in which dislocation lines move and interact with each other according to physics-based dynamical rules.\cite{KUBIN-08}  This computational scheme has intrinsic limitations \cite{LESAR-14}, and so far has not been able to describe anything beyond the earliest stage of strain hardening.  However, this situation is changing with the advent of true molecular dynamics simulations of realistic three-dimensional crystals subject to shear stresses.  Recent results by Zepeda-Ruiz {\it et al} \cite{Bulatov-17}, using one of the world's most powerful computers at Livermore National Laboratory, have proven to be enormously important in testing the theory that I will describe in the rest of this paper.\cite{JSL-18}

\section{Thermodynamic Dislocation Theory: The Effective Temperature} 

My plan for this paper is to present just the simplest possible description of the thermodynamic dislocation theory (TDT), focussing on only its main features and briefly describing some of the results.  More technical details can be found in the previously published papers, especially \cite{JSL-18,LBL-10,JSL-15,JSL-17rev}.  

There are two main features that distinguish the TDT from earlier descriptions of dislocation dynamics.  These are, first, the use of an effective disorder temperature for describing nonequilibrium behavior consistent with the second law of thermodynamics. Second is a depinning analysis that, in a simple way, describes the behavior of Cottrell's birds' nest.  I start with the effective temperature.

In contrast to amorphous plasticity \cite{FL-11}, where identifying shear transformation zones or the like has always been somewhat problematic, the elementary flow defects in crystals -- the dislocations -- are unambiguous.  They are easily identifiable line defects, whose dynamic time scales are longer than those of the ambient thermal fluctuations by many orders of magnitude.  They have well defined energies and easily visible configurations.  However, as stated above, they are intrinsically nonequilibrium entities.  They are the agents of deformation and dissipation when external forces drive energy to flow through the system.  Under the influence of these forces, the dislocations undergo complex chaotic motions, so that it becomes both possible and necessary to describe their behavior statistically.  That statistical analysis is literally thermodynamic.  

To explore this picture, it is useful to start by thinking of a slab of material lying in the plane of an applied shear stress, undergoing only uniform, steady-state deformation.  Then focus only on the dislocations.  That is, assume that, because of their very large energies and slow time scales, the dislocations are almost -- but not completely -- decoupled from all the other kinetic and vibrational degrees of freedom in this system. The dislocation lines oriented perpendicular to this plane are driven by the  stress to move through the system, producing shear flow. Let the area of this slab be $A_0$ and, for the sake of argument, let its thickness be a characteristic dislocation length, say $L_0$.  Denote the configurational energy and entropy of the dislocations in this slab by $U_0(\rho)$ and $S_0(\rho)$ respectively, where, $\rho$ is the areal density of dislocations or, equivalently, the total length of dislocation lines per unit volume. The entropy $S_0(\rho)$ is computed by counting the number of arrangements of dislocations at fixed values of $U_0$ and $\rho$.  

The dislocations are driven by the applied stress to undergo motions that are chaotic on deformation time scales; that is, they explore  statistically significant parts of their configuration space.  According to Gibbs, this configurational subsystem maximizes its entropy; that is, it moves toward states of maximum probability.  It does this at a value of the energy $U_0$ that is determined by the balance between the input power and the rate at which energy is dissipated into a thermal reservoir.  The method of Lagrange multipliers tells us to find this most probable state by maximizing the function $S_0 - (1/\chi)\, U_0$, and then finding the value of the multiplier $1/\chi$ for which $U_0$ has the desired value.  Thus $ \chi \equiv k_B T_{e\!f\!f}$; and the free energy to be minimized is
\begin{equation}
\label{Fdef}
F_0 = U_0 - \chi\,S_0. 
\end{equation}

Minimizing $F_0$ in Eq.~(\ref{Fdef}) determines the steady-state dislocation density, say  $\rho_{ss}$, as a function of the steady-state effective temperature $\chi_{ss}$. In the simplest approximation, $U_0 = A_0\,e_D\, \rho$, where $e_D$ is a characteristic energy of a dislocation of length $L_0$.  Similarly, we can estimate the $\rho$ dependence of the entropy $S_0$ by dividing the  area $A_0$ into elementary squares of area $a^2$, where $a$ is the minimum spacing between noninteracting dislocations -- an atomic length scale, somewhat larger than the length of the Burgers vector $b$.  Then we count the number of ways in which we can distribute $\rho\,A_0$ line-like dislocations, oriented  perpendicular to the plane, among those squares. The result has the familiar form $S_0 = -\,A_0\,\rho\,\ln(a^2\,\rho) + A_0\,\rho$.  Minimizing $F_0$ with respect to $\rho$ produces the usual Boltzmann formula, 
\begin{equation}
\label{rhoss}
\rho_{ss} = {1\over a^2}\,e^{-\,e_D/\chi_{ss}}. 
\end{equation}
We see that an appreciable density of dislocations requires a value of $ \chi_{ss}$ that is comparable to $e_D$, which is enormously larger than the ambient thermal energy $k_B\,T$.   

Next, note that $\chi$ is a measure of the configurational disorder in the material, in direct analogy to the way in which the ambient temperature $T$ determines the intensity of low-energy fluctuations.  As such, $\chi_{ss}$ must be a function primarily of the plastic strain rate $\dot\epsilon^{pl}$, which determines the rate at which the atoms and dislocations are being caused to undergo rearrangements.  If this strain rate is slow enough that the system has time to relax between rearrangement events, then the steady state of disorder is determined only by the number of rearrangements that have occurred and not by the rate at which they occurred.  

This argument means that $\chi_{ss}$ must be some nonzero constant, say $\chi_0$, at strain rates appreciably smaller than atomic vibration frequencies; that is, roughly, $\dot\epsilon^{pl} \le 10^6\,s^{-1}$, which is true for all but strong-shock experiments. We can even make a rough estimate of $\chi_0$ by guessing (in the spirit of Lindemann's melting criterion) that the transition to a rate-dependent $\chi_{ss}$ occurs when the average spacing between dislocations is about ten times the minimum spacing $a$; i.e., from Eq.(\ref{rhoss}), $e_D/\chi_0 \sim 2\,\ln(10) \sim 4$.  The resulting value $\chi_0/e_D \sim 0.25$ is quite close to what is found experimentally; it is the value that I have used in all the experimental comparisons shown in Sec.\ref{Expt}. 

It follows from Eq.(\ref{rhoss}) that $\rho_{ss}$ is independent of strain rate under essentially all steady-state experimental conditions.  As will be seen in the next Section, the driving stress is determined primarily by the dislocation density, and therefore must also be nearly independent of strain rate. In fact, the steady-state stress for room-temperature copper increases by less than a factor of $2$ between strain rates of $10^{-3}\,s^{-1}$ and $10^8\,s^{-1}$.  I find it remarkable that this previously unexplained, qualitative feature of the experimental data can be understood using just the concept of the effective temperature and some simple, dimensional arguments.  

\section{Thermodynamic Dislocation Theory: The Depinning Mechanism}

My proposed solution to Cottrell's birds'-nest problem is to assume that the dominant rate-controlling mechanism during deformation is thermally activated depinning of the entangled dislocations.  The depinning analysis starts with Orowan's dimensional relation between the plastic strain rate $\dot\epsilon^{pl}$, the dislocation density $\rho$, and the average dislocation velocity $v$:
\begin{equation}
\label{Orowan}
\dot\epsilon^{pl}= \rho\,b\,v,
\end{equation}
where $b \sim a$ is the magnitude of the Burgers vector.  If a depinned dislocation segment moves almost instantaneously across the average distance   between pinning sites $\ell = 1/\sqrt{\rho}$ (the average spacing between dislocations), then $v = \ell/\tau_P$, where $1/\tau_P$ is a thermally activated depinning rate given by
\begin{equation}
\label{tauP}
{1\over \tau_P} = {1\over \tau_0}\,e^{- U_P(\sigma)/k_B T}.
\end{equation} 
Here, $\tau_0$ is a microscopic time scale that I usually have chosen to be $10^{-12}s$. 
 
The activation barrier $U_P(\sigma)$ must be a decreasing function of the stress $\sigma$. For dimensional reasons, $\sigma$ should be expressed here in units of some physically relevant stress, which we can identify as the Taylor stress $\sigma_T$.  Suppose that a pinned pair of dislocations must be separated by a distance $a' \ll a$ in order to break the bond between them. If these dislocations remain pinned to other dislocations at distances $\ell$,  then this displacement is equivalent to a strain of order $a'/\ell= a'\sqrt{\rho}$ and a corresponding stress of order $\mu\,a'\,\sqrt{\rho}$, where $\mu$ is the shear modulus.   Thus  
\begin{equation}
\label{sigmaT}
\sigma_T (\rho)= \mu\,{a'\over \ell} \equiv \mu_T\,\sqrt{a^2\,\rho};~~~\mu_T = (a'/a)\,\mu,
\end{equation}
where $\sigma_T$ is the Taylor stress, rederived here by an argument somewhat different from the one that Taylor used in his 1934 paper.\cite{TAYLOR-34} The quantity $\mu_T$ may also contain a dimensionless factor of order unity to correct for uncertainty in the exponential function assumed in the following equation for $U_P(\sigma)$.  As in \cite{LBL-10,JSL-17rev}, write
\begin{equation}
\label{UP}
U_P(\sigma) = k_B\,T_P\,e^{- \sigma/\sigma_T(\rho)},
\end{equation}
where $k_B\,T_P$ is the pinning energy at zero stress. The exponential function used here has no special significance; it is just the simplest decreasing function of  $\sigma/\sigma_T$ that neither vanishes nor diverges at finite values of its agument.

The Orowan formula for the strain rate in Eq.(\ref{Orowan}) becomes 
\begin{equation}
\label{qdef}
\dot\epsilon^{pl} = {b\over \tau_0}\,\sqrt{\rho}\, \exp \Bigl[- {T_P\over T} e^{-\sigma/\sigma_T(\rho)}\Bigr].
\end{equation}
Now solve Eq.(\ref{qdef}) for $\sigma$ as a function of $\rho$, $q\equiv \tau_0\,\dot\epsilon^{pl}$, and  $T$. The result is:
\begin{equation}
\label{sigmadef}
\sigma = \sigma_T(\rho)\,\,\nu(\rho,q,T),
\end{equation} 
where 
\begin{equation}
\label{nudef}
\nu(\rho,q,T) = \ln\Bigl({T_P\over T}\Bigr) - \ln\Biggl[\ln\Bigl({b\sqrt{\rho}\over q}\Bigr)\Biggr] .
\end{equation}
Note that $\nu$ is a very slowly varying function of its arguments, consistent with the well known but approximate validity of the Taylor formula in Eq.(\ref{sigmaT}). This result is also consistent with the observation at the end of the preceding Section that, if $\rho$ is independent of strain rate, then the steady-state stress must also be very nearly a constant.  The converse of this observation is that the strain rate given by the double-exponential formula in Eq.(\ref{qdef}) is an extremely rapidly varying function of the stress and the temperature.  As will be seen below, this solution of the birds'-nest problem solves other long-standing puzzles about yielding transitions, banding instabilities and the like. 

\section{Equations of Motion}
\label{EOM}

The discussion so far has pertained primarily to steady-state deformations.  To address issues such as strain hardening, however, we need time dependent equations of motion.  In what follows, I talk primarily about spatially uniform situations, although I will mention one example in which a spatial nonuniformity is important.  

Start by assuming that the elastic and plastic shear rates are  additive, i.e. $\dot\epsilon^{total} = \dot\epsilon^{el} + \dot\epsilon^{pl}$. Then the equation of motion for the stress $\sigma$ is
\begin{equation}
\label{dotsigma}
\dot\sigma = \mu\,\dot\epsilon^{el}= \mu\,(\dot\epsilon^{total} - \dot\epsilon^{pl}),
\end{equation}
where $\mu$ is the shear modulus.  The crucial ingredient here is $\dot\epsilon^{pl}$, which is given in Eq.(\ref{qdef}) as a function of the dynamical variables $\sigma$, $\rho$, and the temperature $T$. Thus, this equation, like the others that follow, is highly nonlinear.

The equation of motion for the dislocation density $\rho$ is:
\begin{equation}
\label{dotrho} 
\dot\rho = \kappa_{\rho}\,{\sigma\,\dot\epsilon^{pl}\over \gamma_D}\,\Bigl[1- {\rho\over \rho_{ss}(\chi)}\Bigr],
\end{equation}
where $\gamma_D \sim e_D/ L_0$ is the dislocation energy per unit length, and $\kappa_{\rho}$ is the fraction of the input power $\sigma\,\dot\epsilon^{pl}$   that is converted into dislocations.  The second term inside the square brackets in Eq.(\ref{dotrho}) determines the rate at which dislocations are annihilated.  It does this by invoking a detailed-balance approximation using the effective temperature $\chi$; that is, it says that the density $\rho$ must approach the value given by Eq.(\ref{rhoss}), but with the steady-state $\chi_{ss}$ replaced by the time dependent $\chi$ during the approach to steady state deformation.  

Note that Eq.(\ref{dotrho}) describes the flow of energy in and out of the subsystem of the dislocations.  It uses the effective temperature in an essential way; we would not have been able to write this equation without that thermodynamic basis for the theory.  With it, however, we do not need detailed information about the mechanisms by which the dislocations are annihilated; we simply need to require that those mechanisms be consistent with the second law of thermodynamics.  All of the detailed physical ingredients of this equation are contained in the conversion factor $\kappa_{\rho}$, which describes dislocation creation.  But now the flow of information is reversed in comparison with what happened with the phenomenological curve-fitting procedures.  The general structure of Eq.(\ref{dotrho}) is not controversial; it is based on well understood physical principles.  So now, by measuring the dependence of $\kappa_{\rho}$ on quantities such as strain rate or grain size or the like, we learn new physics.  

\begin{figure}[h]
\begin{center}
\includegraphics[width=\linewidth] {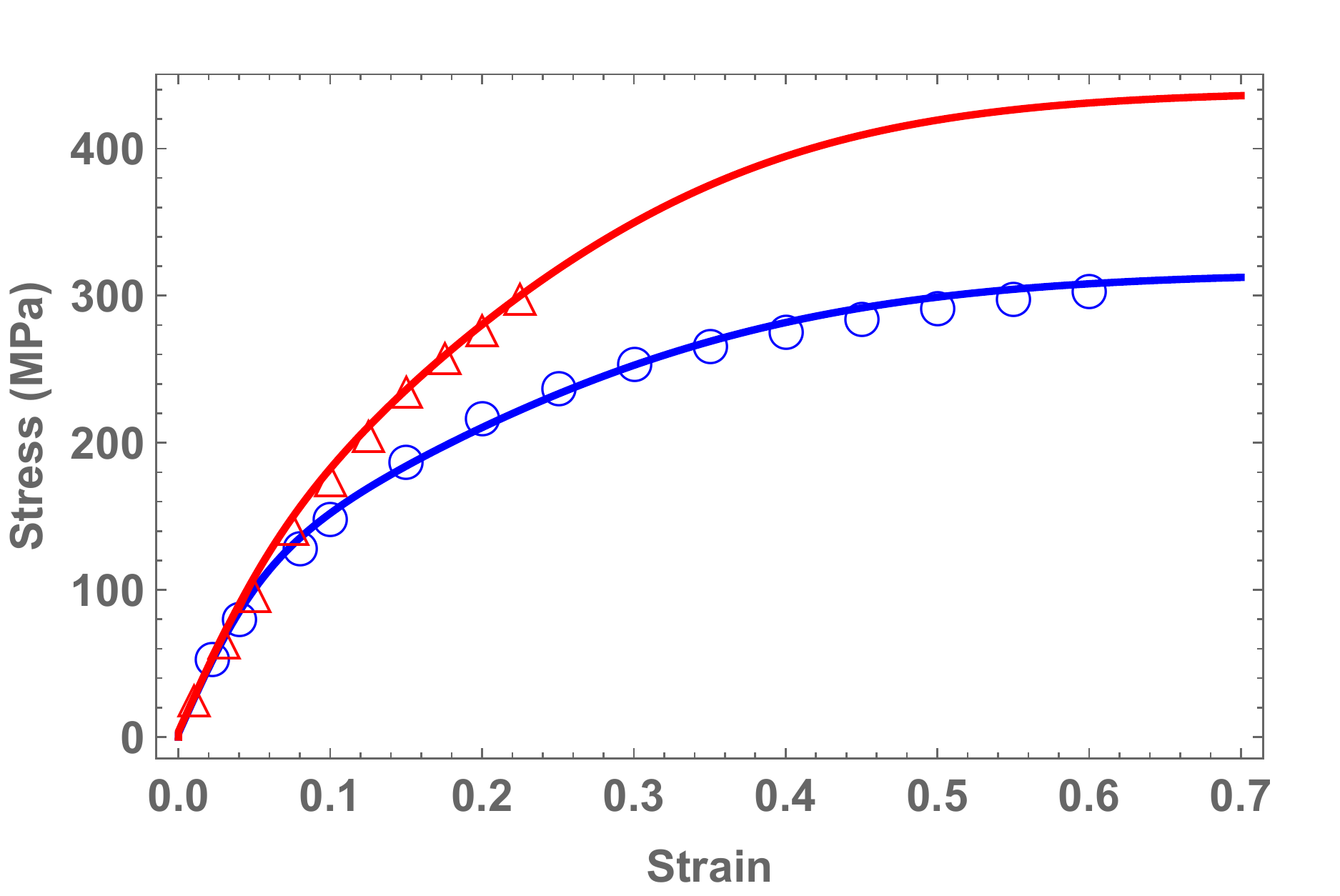}
\caption{Experimental data and theoretical stress-strain curves for copper at $T=298\,K$, for strain rates $0.002\,s^{-1}$ (lower blue curve) and $2,000\,s^{-1}$ (upper red curve).}   \label{PCH-Fig-1}
 \end{center}
\end{figure}

The equation of motion for $\chi$ is a statement of the first law of thermodynamics, which can be written in the form \cite{JSL-17rev}
\begin{equation}
\label{dotchi}
c_{e\!f\!f}\,\dot\chi = \sigma\,\dot\epsilon^{pl}\,\Bigl( 1- {\chi\over\chi_{ss}}\Bigr) - \gamma_D\,\dot\rho.
\end{equation}\\
Here, $c_{e\!f\!f}$ is the effective specific heat, given by $V\,c_{e\!f\!f} = \chi\,\partial S/\partial\chi$, with $V$ being the volume. The second term in the parentheses is proportional to the rate at which effective heat is converted to ordinary heat; the factor $\chi$ is actually $\chi - k_BT$ but with $k_BT\ll \chi$.  The last term in this equation is the rate of energy storage in the form of dislocations.  

Finally, because $\dot\epsilon^{pl}$ in Eq.(\ref{qdef}) is such a rapidly varying function of $T$, we need an equation of motion for the ordinary temperature.  This is simply
\begin{equation}
\label{dotT}
c_T \dot T = \beta\,\sigma\,\dot\epsilon^{pl} - {\cal K}_0\,(T - T_0) + {\cal K}_1\,\nabla^2 T,
\end{equation}
where $c_T$ is the ordinary thermal specific heat, $\beta$ is the Taylor-Quinney factor that determines what fraction of the input power is converted directly into heat, $T_0$ is the ambient temperature, and ${\cal K}_0$ and ${\cal K}_1$ are thermal transport coefficients.  

\section{Comparisons with Experiment}
\label{Expt}

Solutions of Eqs.(\ref{dotsigma} - \ref{dotT}) and comparisons with experimental data have been published in Refs.\cite{LBL-10,JSL-17rev,JSL-17,LTL-17,LTL-18}.  Here I  summarize only a few of those results to illustrate points made in the preceding discussion.  More details, including all parameter values for these selected cases, can be found in Ref.\cite{JSL-17rev}.  

\begin{figure}[h]
\begin{center}
\includegraphics[width=\linewidth] {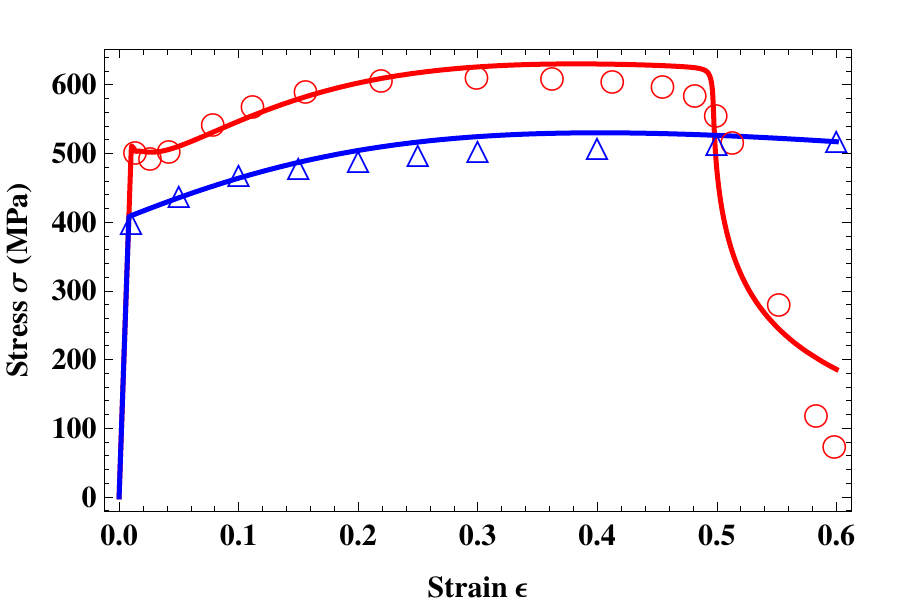}
\caption{Experimental data and theoretical stress-strain curves for steel at (nominally) $T=300\,K$, for strain rates $0.0001\,s^{-1}$ (lower blue curve) and $3,300\,s^{-1}$ (upper red curve). Both curves show sharp yielding transitions at small strain.  The upper curve shows the onset of adiabatic shear banding at a strain roughly equal to $0.5$. }   \label{PCH-Fig-2}
 \end{center}
\end{figure}

Start with the strain-hardening curves for room-temperature copper that are shown in Fig. \ref{PCH-Fig-1} for two very different strain rates, $0.002\,s^{-1}$ and $2,000\,s^{-1}$.  These experimental curves were taken from \cite{LANL-99}.  The theoretical curves are solutions of the equations of motion for constant strain rates.  They use two system-specific physical parameters, $T_P = 40,800\,K$ and $\mu/\mu_T = 31$, which should be independent of both strain rate and temperature.  The other parameters needed for plotting these curves are $\chi_0/e_D = 0.25$ plus the constant factors $\kappa_{\rho}$ and $c_{e\!f\!f}$, and initial values of $\rho$ and $\chi$, all of which were assumed to be the same for both the high- and low-strain-rate cases.  There are no arbitrary power laws or assumptions about transitions between various ``stages'' of hardening.  The physical mechanisms that determine the shapes of these curves are contained in the conversion factors $\kappa_{\rho}$ and $c_{e\!f\!f}$. 

The conversion factor $\kappa_{\rho}$ is especially interesting.  Kocks and Mecking \cite{KOCKS-MECKING-03} discovered experimentally that the onset slope of these curves seemed to be a constant, independent of both strain rate and temperature.  Apparently, the initial values of $\rho$ for these copper samples were much smaller than their steady-state values, so that there is no apparent yield stress, and the second term in the brackets on the right-hand side of Eq.(\ref{dotrho}) can be neglected at small strains.  Then a simple calculation tells us that
\begin{equation}
\label{onset}
{1\over \mu}\,\Bigl({\partial\sigma\over \partial\epsilon}\Bigr)_0 \cong \kappa_{\rho}\,{b^2\,\mu_T^2\,\nu_0^2\over 2\,\mu\,\gamma_D},
\end{equation}
where the subscript $0$ denotes the onset value at small strain.  $\nu_0$ is the onset value of the slowly varying function $\nu$ given in Eq.(\ref{nudef}).  Thus, the strain rate has effectively cancelled out of this formula.  Moreover, since $\mu$, $\mu_T$, and $\gamma_D/b^2$ (each with dimensions of energy per unit volume) should all scale with temperature in about the same way, the right-hand side of Eq.(\ref{onset}) should be independent of temperature.  Thus, the TDT has explained the observation of Kocks and Mecking, assuming that $\kappa_{\rho}$ remains constant.

Even more interestingly, the physical interpretation of $\kappa_{\rho}$ as an energy conversion factor tells us that it cannot generally remain constant, but must contain information about dislocation-creation mechanisms.  For example, the data of Meyers et al \cite{MEYERSetal-95}, as interpreted in \cite{JSL-17}, reveals that $\kappa_{\rho}$ increases with the inverse square root of decreasing grain size.  In other words, $\kappa_{\rho}$ contains a term  proportional to the stress-concentration factor near a corner of a typical grain,  so that smaller grains are more effective sources of new dislocations.  Then, using this conversion term to compute dislocation densities via Eq.(\ref{dotrho}) and, ultimately, flow stresses, we find a simple explanation of Hall-Petch grain-size effects -- far more compelling, in my opinion, than the conventional way of attributing these effects to pile-ups of dislocations at grain boundaries and using the dubious stress-additivity assumption.\cite{ARMSTRONG-HP14}  

Turn now to Fig.\ref{PCH-Fig-2}.  The data points here are taken from the classic 1988 study of adiabatic shear banding in steel by Marchand and Duffy.\cite{MARCHAND-DUFFY-88}   The theoretical curves are from \cite{JSL-17rev}.   See  \cite{LTL-18}  for a theoretical analysis of all the data in \cite{MARCHAND-DUFFY-88} and for graphs of space- and time-dependent strain rates and temperatures; and see \cite{JSL-17rev} for all of the parameter values used computing these particular curves. 

The upper red stress-strain curve in Fig.\ref{PCH-Fig-2} is measured at a high strain rate, $3,300\,s^{-1}$; the lower blue curve is effectively quasistatic, $0.0001\,s^{-1}$.  Both curves are measured nominally at room temperature.  By ``nominally,'' I mean that the measurements were made on samples that initially were at room temperature, but that, as will be seen, internal heating effects were important at the high strain rate.  For both curves, $T_P = 6\times 10^5\,K$, $\mu = 5\times 10^4\,MPa$, $\mu_T= 1200\,MPa$, and $\chi_0/e_D = 0.25$.   The thermal transport coefficients ${\cal K}_0$ and ${\cal K}_1$ have both been set to zero. 

Both of the stress-strain curves in Fig.\ref{PCH-Fig-2} exhibit sharp transitions between elastic and plastic deformation at small strains of order $0.02$.  These yielding transitions are predicted by the TDT with no assumptions other than those already stated; they result from the strong stress-sensitivity of the strain rate predicted by  Eqs.(\ref{qdef} - \ref{nudef}).  The yield stresses are not fitting parameters.  In plotting these curves, I have assumed that the initial dislocation density was approximately the same for both, and that the principal difference between the yield stresses was caused by the weak but non-negligible rate dependence of the function $\nu$ in Eq.(\ref{nudef}). Thus, the quantitative agreement between theory and experiment shown here is a nontrivial test of the theory.  (The small overshoot at the yield point for the upper curve is most probably an instrumental artifact.  I reproduced it artificially here by adjusting the initial effective temperature, in effect, assuming that the overshoot was caused by sample preparation.) 

The abrupt stress drop at $\epsilon \cong 0.5$ on the fast curve in Fig.\ref{PCH-Fig-2} indicates the onset of an adiabatic shear-banding instability.  This instability was triggered by a narrow scratch inscribed on the surface of the sheared sample.  ``Adiabaticity''  refers to the fact that the instability is caused by thermal softening in a situation where heat flow is slower than plastic deformation.  A local increase in strain rate at the triggering defect produces a local increase in heat generation {\it via} the first term on the right-hand side of Eq.(\ref{dotT}).  In turn, this increase in temperature increases the local strain rate according to  Eq.({\ref{qdef}), which further increases heat generation.  The result is a runaway instability if heat is unable to flow away from the hot spot more quickly than new heat is generated there.  This thermal conduction is described by the last two terms on the right-hand side of Eq.(\ref{dotT}), which both have been set to zero in this example.  Thus, we are looking at a delicate balance between thermal and mechanical behaviors that, in this case, is governed primarily by the strong temperature sensitivity of the depinning mechanism.  The experimentally observed stress drop is sharper than the theoretical one because  shear banding almost certainly changes into fracture in its late stages.  

My choices of experimental examples in the preceding paragraphs are intended to demonstrate that the TDT is, indeed, ``predictive'' in the sense that I think was meant by Cottrell.  The equations of motion in Sec.\ref{EOM} are based entirely on general fundamental principles -- the laws of thermodynamics, energy conservation, and dimensional analysis.  Specific phenomena such as hardening rates, grain-size effects or yielding transitions played no role in writing them down.  Those phenomena were {\it predicted} by the equations, and these predictions tell us that the basic assumptions, if not exactly correct or complete, must be on the right track for further exploration.\\

\begin{acknowledgments}

JSL was supported in part by the U.S. Department of Energy, Office of Basic Energy Sciences, Materials Science and Engineering Division, DE-AC05-00OR-22725, through a subcontract from Oak Ridge National Laboratory.  

\end{acknowledgments}

\end{document}